\documentclass{emulateapj}

\usepackage{amssymb,amsmath,bm}
\usepackage{graphicx}
\usepackage[colorlinks]{hyperref}
\usepackage{natbib}
\usepackage{graphicx}
\usepackage{subfigure}

\usepackage{xcolor}
\usepackage[normalem]{ulem}

\begin{document}

\title{Simultaneous chiral symmetry restoration and deconfinement \\ Consequences  for the QCD phase diagram}

\author{Thomas Kl{\"a}hn$^{1,*}$, Tobias Fischer$^1$, and  Matthias Hempel$^2$}
\affil{$^1$ Institute of Theoretical Physics, University of Wroclaw, Pl. M. Borna 9, 50-204 Wroclaw, Poland\\
$^2$ Department of Physics, University of Basel, Klingelbergstrasse 82, 4056 Basel, Switzerland\\
$^*$ thomas.klaehn@ift.uni.wroc.pl}

\begin{abstract}
For studies of quark matter in astrophysical scenarios the thermodynamic bag model (tdBag) is commonly employed. Although successful, it does not account for dynamical chiral symmetry breaking (D$\chi$SB) and repulsions due to the vector interaction
which is crucial to explain recent observations of massive, two solar mass neutron stars. In \citet{Klaehn:2015} we developed the novel vBag quark matter model which takes these effects into account.  This article extends vBag to finite temperatures and isospin asymmetry.  Another particular feature of vBag is the determination of the deconfinement bag constant $B_{\rm dc}$  from a given hadronic equation of state (EoS) in order to ensure that chiral and deconfinement transitions coincide. We discuss consequences of this novel approach for the phase transition construction, the phase diagram, and implications for protoneutron stars.
\end{abstract}

\date{\today}


\maketitle
\bibliographystyle{apj}

\section{Introduction}
\label{intro}

The theory of strong interactions, i.e. Quantum Chromodynamics (QCD), considers hadrons and mesons as color neutral compound objects of quarks and gluons. 
Besides confinement a second key feature of QCD is dynamical chiral symmetry breaking (D$\chi$SB) and its restoration at large densities and high temperatures. 
Currently, lattice-QCD is the only ab-initio approach to solve QCD numerically~\citep[cf.][and references therein]{Fodor:2004nz,Aoki:2006we}. 
In the vicinity of vanishing chemical potentials lattice-QCD predicts a cross-over phase transition for both deconfinement and chiral symmetry breaking in the temperature range $150<T<170$~MeV~\citep[the currently considered value is $154\pm9$~MeV, cf.][]{Katz:2012JHEP,Laermann:2012PRD,Laermann:2012PRL,Bazavov:2014pvz,Katz:2014PhLB}.
This predicted temperature is in agreement with hadron freeze out data from heavy-ion collision experiments~\citep[cf.][and references therein]{Redlich:2015}.
However, lattice-QCD calculations are restricted to low chemical potentials.
On the other hand, in astrophysical systems, e.g., the interior of neutron stars and core collapse supernovae, we usually encounter large chemical potentials (or equivalently high densities even in excess of normal nuclear density), high isospin asymmetries, and finite temperatures. 
Such conditions are inaccessible by lattice-QCD and the current generation of heavy-ion collision experiments.

The two most commonly used effective quark matter models in astrophysics are the thermodynamic bag model (tdBag) of \citet{Farhi:1984qu} and
models of the Nambu-Jona-Lasinio (NJL) type~\citep[cf.][]{Nambu:1961tp,Klevansky:1992qe,Buballa:2003qv}. 
Recently, we illustrated in \citet{Klaehn:2015} that both approaches can be understood as limiting solutions of QCD's in-medium Dyson-Schwinger gap equations within a particular set of approximations.
The advantage of the Dyson-Schwinger formalism is a closer connection to QCD in comparison to the aforementioned phenomenological approaches.
First explorations of the EoS of deconfined quark matter within this approach have been performed~\citep[cf.][]{Chen:2008zr,Klahn:2009mb,Chen:2011my,Chen:2015mda,Chen:2016ran}. 
Perturbative QCD is valid in the limit of asymptotic freedom where quarks are no longer strongly coupled~\citep[cf.][and references therein]{Kurkela:2014vha}. 
Therefore, it provides a valuable benchmark for the asymptotic behaviour of realistic quark matter model EoS.

Currently, no consistent approach exists to describe medium properties of hadrons and mesons at the level of quarks and gluons, in particular at high density. 
Hence, the deconfinement phase transition is usually constructed from a given hadronic EoS with hadrons and mesons as the fundamental degrees of freedom
and an independently computed quark matter EoS.  Only few attempts exist to improve this situation, see, e.g., \citet{Dexheimer:2009hi,Steinheimer:2010ib}.
Constructions based on Maxwell's conditions by definition result in a 1st-order phase transition~\citep[cf.][]{Rischke:1988,Barz:1989}. 
In \citet{Fischer:2011} a  Gibbs construction based on tdBag as the quark matter EoS at finite temperatures has been employed for applications in astrophysics. 
\citet{Yasutake:2014} considered pasta-structures at the quark-hadron interface and compared the results with standard Maxwell and Gibbs constructions. 
Further alternative approaches for the construction of possible quark-hadron phase transitions were discussed in \citet{Hatsuda:2013} and in \citet{Baym:2015}.

In \citet{Klaehn:2015} we introduced the novel quark matter EoS vBag in particular for applications in astrophysics. 
Although vBag is initially motivated as a simplification of a NJL model with scalar and iso-vector condensates
it is essentially an extended bag model which explicitly accounts for the repulsive vector interaction.
This property is essential for a stiffening of the EoS towards high densities and results in maximum neutron star masses in agreement with the current constraint of about 2~M$_\odot$ from the observations of \citet{Antoniadis:2013} and \citet{Demorest:2010}, the latter recently revised to $1.928\pm0.017$~M$_\odot$ by \citet{Fonseca:2016tux}.
Moreover, vBag mimics (de)confinement via a phenomenological pressure correction to the EoS in form of a deconfinement bag constant $B_{\rm dc}$.
The simultaneous occurrence of chiral symmetry restoration and (de)confinement is generally not provided for a `standard' Maxwell or Gibbs construction. This represents an essential assumption for this study. The consequences will be discussed in detail. The idea to enforce deconfinement at the chiral transition by manually equating the pressure of hadronic and quark matter EoS at zero temperature has been discussed previously by \cite{Pagliara:2007ph}.

In this article we extend vBag to finite temperatures and isospin asymmetry. 
As a consequence of our model prescription, the deconfinement bag constant $B_{\rm dc}$ becomes medium dependent and dependends directly on the hadronic EoS. The medium dependence induces correction terms for the EOS of the quark phase, which consequently depends on the hadronic EOS as well.
For illustration we select a nuclear EoS from the catalogue of \citet{Hempel:2009mc}. 
It treats the nuclear medium within a quasi-particle picture of the relativistic mean-field (RMF) framework for conditions in excess of normal nuclear matter density and high temperatures. Specifically, we choose the parametrization DD2 from \citet{Typel:2009sy}. 
We aim to make vBag accessible to a broader spectrum of astrophysics applications where the appearance of quark matter can be studied, e.g., simulations of neutron star mergers, core collapse supernovae and protoneutron star cooling~\cite[cf.][]{Pons:2001ar,Sagert:2008ka,Nakazato:2008su,Fischer:2011}.

In Sec.~\ref{vbag} we review the basic equations of vBag including novel aspects due to the extension to finite temperature and isospin asymmetry. 
In Sec.~\ref{phase} we construct the hybrid EoS HS(DD2)--vBag based on 
a quark matter EoS parametrization that was already used in \citet{Klaehn:2015}. 
We describe the resulting phase diagram and differences
to the standard approach for the construction of the phase transition. 
In Sec.~\ref{pns} we discuss a possible astrophysical applications at the example of protoneutron stars. Sec.~\ref{limits} is devoted to limitations of this study with a detailed comparison of vBag and NJL.
The manuscript closes with the summary in Sec.~\ref{summary}.

\section{vBag as a hybrid approach}
\label{vbag}
\subsection{Chiral symmetry and vector interaction}
In \citet{Klaehn:2015} we introduced the vBag model as an extension of tdBag, which in turn was introduced in \citet{Farhi:1984qu}.
As typical for quasi-particle approaches the thermodynamical behavior of vBag is fully expressed in
terms of non-interacting Fermi gas integrals.
The model considers only the chirally restored phase where the quark mass is well approximated
by the bare quark mass. 
Despite this similarity to the standard bag model vBag accounts explicitly for the repulsive vector interaction and 
the dynamical breaking of chiral symmetry at a critical chemical potential $\mu_{\chi,f}$,
where $f$ denotes the quark flavor.

For $\mu_f<\mu_{\chi,f}$ 
chiral symmetry is broken and we assume that quarks are confined in 
hadron and meson states which are not accessible for vBag.
The chiral phase transition and the corresponding critical chemical potential $\mu_{\chi,f}$ is defined 
by the value of the chiral bag constant $B_\chi$, demanding that  at $\mu_f = \mu_{\chi,f}$ the pressure of a single
quark flavor, that is given as
\begin{eqnarray}
\label{eq:P1f}
P_f(\mu_f) &=& P_{{\rm FG},f}^{kin}(\mu_f^*) + \frac{K_v}{2}  n_{{\rm FG},f}^2(\mu_f^*) - B_{\chi,f}~ \; ,
\end{eqnarray}
turns positive (for illustrations cf. Fig.~(1) of \citet{Klaehn:2015}).  
$B_{\chi,f}$ is a flavor dependent parameter which can be computed from a microscopic approach 
as the difference between the vacuum pressure of the chirally broken and the chirally restored phase.

As this prescription is applied for each flavor independently
it would result in sequential chiral transitions of light and heavy quarks \citep[more details are given in ][]{Klaehn:2015}. 
We return to this point at the discussion of the phase transition in Sec.~\ref{SEC:DEC}. 

The constant $K_v$ in the second term in Eq.~(\ref{eq:P1f}) relates directly to the coupling strength $G_v$
of the vector current-current interaction as one would define it in NJL type models.
As explained in more detail in \citet{Klaehn:2015}, $B_{\chi,f}$  (besides the regularization scheme and involved parameters)
depends on a corresponding scalar coupling constant $G_s$ which itself relates to $G_v$ by $G_v=G_s/2$.
Hence, $K_v$ and $B_{\chi,f}$ are not strictly independent quantities.
However, as the latter relation is not expected to hold strictly in a realistic model
we consider these two as rather independent model parameters of vBag.
A consequence of accounting for the vector interaction which leads beyond the
tdBag model is the appearance of the effective chemical potential $\mu_f^*$
which enters all Fermi gas expressions.
The actual chemical potential $\mu_f$ in the sense of a thermodynamic variable
is determined {\it post-priori} as
\begin{eqnarray}
\label{eq:mugap}
\mu_f &=& \mu_f^*+K_v n_{{\rm FG},f}(\mu_f^*)~.
\end{eqnarray}
This again is common for quasi-particle models where interactions
dynamically alter particle in-medium properties, e.g., mass and chemical potential but not
the structure of the Fermi integrals.

With Eqs.~(\ref{eq:P1f}) and (\ref{eq:mugap}) the EoS at zero temperature
is given and the two relations can be used to determine the energy density and quark number density
of a given flavor,
\begin{eqnarray}
\label{eq:e1f}
\varepsilon_f(\mu_f) &=& \varepsilon_{{\rm FG},f}^{kin}(\mu_f^*) +
 \frac{K_v}{2}  n_{{\rm FG},f}^2(\mu_f^*) + B_{\chi,f}~, \\
n_f(\mu_f) &=& n_{{\rm FG},f}(\mu_f^*)~.
\end{eqnarray}
To extend the model to finite temperatures we
replace the zero temperature Fermi-gas integrals by the equivalent
finite temperature expressions.
The entropy density of the system is expressed in
the quasi-particle picture again,
\begin{eqnarray}
\label{eq:s1f}
s_f(T,\mu_f) &=& \left.\frac{\partial P_f(T,\mu_f)}{\partial T}\right| _{\mu_f}=s_{{\rm FG},f}(T,\mu_f^*).
\end{eqnarray}
One easily verifies the Euler equation $\varepsilon+P=\mu n + T s$ to be fulfilled.
Notice, that in this study we assume a constant, medium independent $B_{\chi,f}$.
This is not a perfect or in general correct choice -- the scalar condensate is known to melt with increasing temperature --
and we assume for the moment not to reach high enough  temperatures to induce 
significant changes of the pressure due to this melting.
We briefly discuss the limitations of this assumption in Sec.~\ref{limits}.

\subsection{Deconfinement}
\label{SEC:DEC}
The previous section described how vBag accounts for the repulsive vector interaction 
and the chiral phase transition of a single quark flavor.
These results can be perfectly understood in terms of a simple NJL model or more general
in the framework of Dyson-Schwinger model studies assuming an underlying gluon contact interaction 
as illustrated in \citet{Klaehn:2015}.
vBag is a limit of these models in the chirally restored phase assuming
that in this domain the bare quark mass is a sufficient approximation of the real mass gap solutions
to describe the thermodynamical behavior of chirally restored quark matter.
The condition $P_f(\mu_{\chi,f})=0$ which determines $\mu_{\chi,f}$ does not imply that the chiral bag constant
describes the pressure under a confining force 
but simply accounts for the fact that
chiral symmetry in confined matter is broken and subsequently the restoration
of chiral symmetry induces pressure to the system.

If one aims to account for both, chiral symmetry restoration and
deconfinement,
it is worth to consider the following two aspects:
{\it First}, a description of hadronic matter that does not take chiral symmetry
breaking on the quark level into account is likely inaccurate.
It is flawed already on the level of the `fundamental' bare hadron masses
independently from the specific modeling of nucleon-nucleon interactions.
{\it Second}, hadronic matter will deconfine after the system reaches 
a sufficient critical chemical potential $\mu_{B, {\rm dc}}$ and 
related pressure.
It is a common approach to describe hybrid nuclear-quark EoS
including the chiral/deconfinement transition without giving
any of these two statements too many thoughts by performing
a Maxwell or Gibbs construction to connect nuclear and quark EoS; 
both EoS can be -- and in astrophysical applications usually are -- 
obtained from different, independent models.

With these considerations in mind we model a hybrid, nuclear--light-quark EoS for applications at finite temperature 
and arbitrary isospin-asymmety.
While an extension to finite temperatures is easily achieved, 
isospin-asymmetry, i.e. $\mu_u\not=\mu_d$ raises the problem
that different light quark chemical potentials imply situations
where, e.g., $\mu_d<\mu_{\chi,(u,d)}<\mu_u$,
and consequently $u$-quarks are chirally restored
while $d$-quarks are not.
This can lead to interesting scenarios~\citep[cf.][]{Blaschke:2008br}.
However, in this work we model an EoS where light flavors appear 
at the same critical baryon chemical potential in analogy to the results obtained with the `standard'
approach for first order phase transitions.
To achieve this we redefine the chiral threshold condition to hold for the total light quark pressure
instead for each single flavor independently,
\begin{eqnarray}
\sum_{f=u,d}P_f(T,\mu_{\chi,f})=0~. \label{eq:P2f}
\end{eqnarray}
Here, $P_f(T,\mu_f)$ is defined as the total quark pressure of a single quark flavor according to Eq.~(\ref{eq:P1f}).
Based on the NJL model parameterization \cite{Grigorian:2006qe} we used in \citet{Klaehn:2015} we assume that the threshold density for the appearance of the heavy strange quark at zero temperature is larger than the light quark threshold. At least in this domain it will not affect the deconfinement bag constant. Possible contributions at finite temperature we omit. To be consistent we neglect hyperon contributions to the hadronic EoS. We point out, that technically our prescription remains the same if one would consider strangeness. This additional degree of freedom is likely to affect the quantitative results of this study but not the prescription we suggest in this work to obtain a medium dependent deconfinement bag constant and the related correction terms to the EoS.

It is useful to introduce the baryon and charge chemical potentials for light quark matter,
which follow in general from $\mu_f=\mathcal{B}_f\mu_B+\mathcal{Q}_f\mu_C$, with $\mathcal{B}_f$ and $\mathcal{Q}_f$ 
being the baryon and electric charge number of the corresponding quark flavor $f$.
This allows us to rewrite
\begin{eqnarray}
\mu_B &=& \mu_u+2\mu_d,\\
\mu_C &=& \mu_u-\mu_d.
\end{eqnarray}
At given $T$ and $\mu_C$ we denote our critical chemical potentials for D$\chi$SB and confinement
as $\mu_{B,\chi}$ and $\mu_{B,{\rm dc}}$ respectively.
The general thermodynamic identities,
\begin{eqnarray}
&&  \frac{\partial P}{\partial \mu_B}=n_B~, \;\;\;\frac{\partial P}{\partial \mu_C}=n_C~, \;\;\;  \frac{\partial P}{\partial T}=s~,
\label{eq:Pder}
\end{eqnarray}
relate the pressure derivatives with respect to the baryonic and charge chemical potentials to the baryon and charge
number densities as well as the temperature derivative to the entropy. 

In this work, we assume thatat sufficiently low temperatures the assumption of a first order phase transition is correct. The condition for the transition from hadronic (H) to quark matter (Q) for the total pressure of each phase reads $P^H(\mu_{B,dc})=P^Q(\mu_{B,dc})$ at given $T$ and $\mu_C$. It defines the deconfinement baryonic chemical potential $\mu_{B,dc}$. We observe: {\it (i)} evidently deconfinement occurs at finite pressure, and  {\it (ii)} without further modifications Eq.~(\ref{eq:P2f}) enforces zero pressure at the chiral transition. Hence, chiral and deconfinement transition are necessarily located at different chemical potentials, $\mu_{B,\chi}<\mu_{B,dc}$. In fact, the gap between these critical chemical potentials easily spans several $100$ MeV. The gap region in this 2-phase approach is obtained by comparing a nuclear matter EoS which does not know about chiral symmetry breaking on the quark level with a quark matter EoS which does not know about confinement. The thermodynamically favored solution in this domain is the hadron EoS due to its higher pressure, and hence the chiral restoration of quark matter is effectively postponed {\it ad hoc} until $\mu_{B,dc}$ is reached\footnote{ Note, how this picture differs from that of the quarkyonic phase~\cite[cf.][]{McLerran:2007qj,McLerran:2009,Blaschke:2010} where chiral symmetry is restored in the confined domain, $\mu_{B,\chi}<\mu_B<\mu_{B,\text{dc}}$. Strictly, this would require to reformulate the nuclear EoS to account dynamically for nucleon masses which depend on medium dependent quark constituent masses.}.
With vBag we could model the same late onset of deconfinement. However, to maintain our assumption of simultaneous chiral symmetry restoration and deconfinement this would require to shift  $\mu_{B,\chi}$ and hence to increase the chiral bag constant to higher values which are not in agreement with, e.g., the  NJL model.

We go the opposite way and want our model to account for simultaneous chiral symmetry breaking and the confinement transition at $\mu_{B,\chi}$. We emphasize that this is an assumption we put into this model and not a general property of vBag.
According to this idea  $\mu_{B,dc}$ has to decrease in order to match $\mu_{B,\chi}$ in agreement with  Eq.~(\ref{eq:P2f}). This is achieved by exploiting that the total pressure is fixed by the previous equations only up to a constant $B_{dc}$,
\begin{equation}
P^Q=\sum_f P_f(T,\mu_f) + B_{\rm dc}.
\end{equation}
Introducing $B_{\rm dc}$ thus enables us to ensure that the pressure of hadronic and quark matter equals at the critical chemical potential $\mu_{B,\chi}$ for the chiral transition. In order to match both pressures one easily finds the value of $B_{\rm dc}=P^H(\mu_{B,\chi})$ at given $T$ and $\mu_C$. In general, we expect $B_{\rm dc}$ to depend on all thermodynamic variables, $T$, $\mu_C$, and $\mu_B$. However, our prescription fixes $B_{\rm dc}$ only on a hyper-surface within this three-dimensional parameter space which leaves one free variable and the corresponding dependence of $B_{\rm dc}$ remains undetermined. We eliminate this degree of freedom by assuming $B_{\rm dc}(T,\mu_C,\mu_B)=B_{\rm dc}(T,\mu_C,\mu_{B,\chi})$, where $\mu_{B,\chi}$ itself is a function of $T$ and $\mu_C$.

\begin{figure*}
\centering
\includegraphics[width=\textwidth]{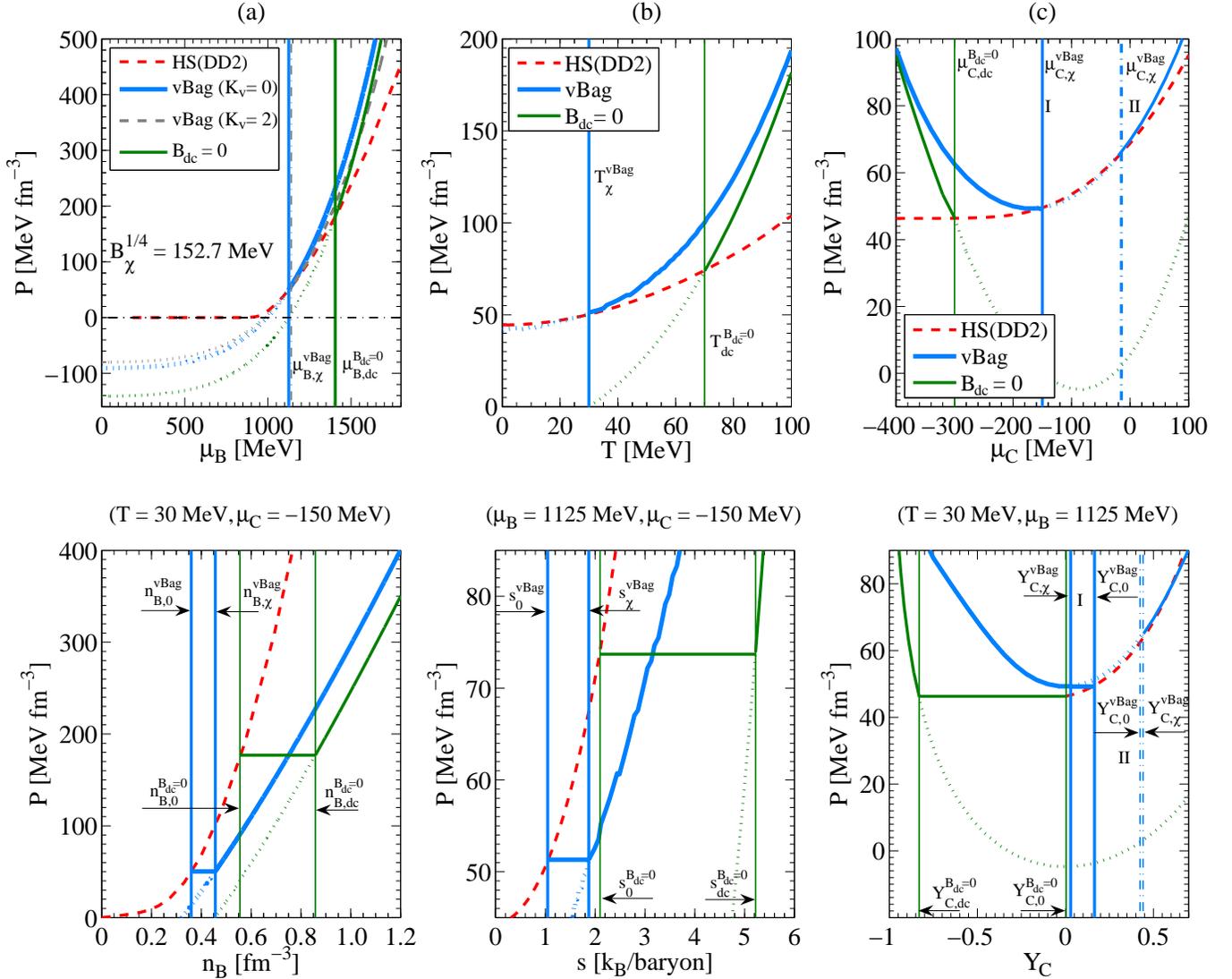}
\caption{(color online) {\em Upper panel:} Phase transition from HS(DD2) to vBag for a selected chiral bag constant $B_\chi$ comparing our and the `standard' approach ($B_{\rm dc}=0$). {\em Top panel:} Dependence on baryochemical potential $\mu_B$ (a), temperature (b), and electronic chemical potential $\mu_C$ (c). {\em Bottom panel:} Dependence of the conjugated thermodynamical variable (see text for definitions). For (a) zero and finite vector coupling $K_v$ (units of $10^{-6}$~MeV$^{-2}$) are considered.}
\label{fig1}
\end{figure*}

\subsection{ Medium Corrections}
We emphasize that we assume that chiral symmetry breaking and deconfinement are tied to the same or close values of the corresponding critical chemical potentials. Whether this will be supported by future research or not, the previous paragraph outlined that both approaches we described imply $\mu_{B,\chi}=\mu_{B, {\rm dc}}$; the traditional phase transition construction of two independent phases by shifting $\mu_{B,\chi}\to\mu_{B, {\rm dc}}$, the $B_{\rm dc}$ approach by shifting $\mu_{B, {\rm dc}}\to\mu_{B,\chi}$. We prefer the latter as it keeps the insights from a chiral description on the quark level and as we show in the following results in an inherent connection of nuclear and quark EoS that goes beyond the `standard' two-phase approach.
Note, that vBag does not require this procedure, and in principle we could obtain any value for $\mu_{B, {\rm dc}}$ with a corresponding choice of $B_{\rm dc}$. In fact, the prescribed procedure can be applied to any quark model which accounts for D$\chi$SB and is not restricted to vBag. However, the problem of an existing domain where the favored nuclear EoS is blind to the restoration of chiral symmetry on the quark level would still stand.

Ignoring the vector interaction induced terms
for the moment, vBag behaves very similar to the thermodynamic bag model as 
it results in an effective bag constant
\begin{eqnarray}
B_\text{eff} = \sum_{f=u,d} B_{\chi,f} - B_\text{dc}~,
\label{eq:Beff}
\end{eqnarray}
which takes similar values as one is used to from the bag model.
We emphasize that unlike for the standard bag model in vBag 
the positive value of $B_\text{eff}$
originates from D$\chi$SB only, while confinement reduces it.
In this sense $B_\text{dc}$ is understood as a
confining binding energy.
Moreover, $B_\text{eff}$ can take values smaller
than in the thermodynamic bag model without
predicting deconfined (light)-quark matter
to be energetically favored over nuclear matter.
This is explained in more detail in \citet{Klaehn:2015}.

As outlined earlier the deconfinement bag constant $B_\text{dc}$
depends on the nuclear EoS.
According to our prescription for the phase transition
it will vary with temperature and asymmetry,  $B_\text{dc}\to B_\text{dc}(T,\mu_C)$,
while we keep it constant with respect to $\mu_B$.
This medium dependence of $ B_\text{dc}$ results in additional contributions
to the charge number and entropy density,
\begin{eqnarray}
&& n_C^Q(T,\mu_C,\mu_B) = \tilde{n}_C^Q(T,\mu_C,\mu_B) + \frac{\partial B_\text{dc}}{\partial \mu_C}~,
\label{eq:nC} \\
&& s^Q(T,\mu_C,\mu_B) = \tilde{s}^Q(T,\mu_C,\mu_B) + \frac{\partial B_\text{dc}}{\partial T}~,
\label{eq:s}
\end{eqnarray}
with $\tilde{n}_C^Q= \sum_{f=u,d} \mathcal{Q}_f n_f = \frac{1}{3}\left(2n_u-n_d\right)$ and $\tilde{s}^Q=\sum_{f=u,d} s_f$.
The derivative terms in \eqref{eq:nC} and \eqref{eq:s} distinguish our model from tdBag and other models with `standard' 
phase transition construction of two independent phases. One obtains
\begin{eqnarray}
&&
\frac{\partial B_\text{dc}}{\partial \mu_C} =: n_{C,\text{dc}}(T,\mu_C) \nonumber \\
&&
=
\frac{\partial}{\partial\mu_C}\left\{P^H\left(T,\mu_C,\mu_B=\mu_{B,\chi}(T,\mu_C)\right)\right\}
\nonumber \\
&&
=
\frac{\partial P^H}{\partial\mu_C}(T,\mu_C,\mu_{B,\chi}) + \frac{\partial\mu_{B,\chi}}{\partial\mu_C}\frac{\partial P^H}{\partial\mu_B}(T,\mu_C,\mu_{B,\chi})
\nonumber \\
&&
=
n_C^H(T,\mu_C,\mu_{B,\chi}) - \tilde{n}_C^Q(T,\mu_C,\mu_{B,\chi}) \frac{n_B^H(T,\mu_C,\mu_{B,\chi})}{n_B^Q(T,\mu_C,\mu_{B,\chi})} \;, \nonumber \\
\label{eq:dBdmuc}
\end{eqnarray}
\begin{eqnarray}
&&
\frac{\partial B_\text{dc}}{\partial T} =: s_\text{dc}(T,\mu_C) \nonumber \\
&&
=
\frac{\partial}{\partial T}\left\{P^H\left(T,\mu_C,\mu_B=\mu_{B,\chi}(T,\mu_C)\right)\right\}
\nonumber \\
&&
=
\frac{\partial P^H}{\partial T}(T,\mu_C,\mu_{B,\chi}) + \frac{\partial\mu_{B,\chi}}{\partial T}\frac{\partial P^H}{\partial\mu_B}(T,\mu_C,\mu_{B,\chi})
\nonumber \\
&&
=
s^H(T,\mu_C,\mu_{B,\chi})
-
\tilde{s}^Q(T,\mu_C,\mu_{B,\chi})\frac{n_B^H(T,\mu_C,\mu_{B,\chi})}{n_B^Q(T,\mu_C,\mu_{B,\chi})}\;, \nonumber \\
\label{eq:dBdT}
\end{eqnarray}
At $\mu_C=0$ one has  isospin symmetric matter
and hence 
$Y_C^H=\tilde{Y}_C^Q=0.5$  (with the charge fractions $Y_C$ defined as $n_C/n_B$), which implies $n_{C,\text{dc}}(T,\mu_C=0)=0$. 
Similary at $T=0$ holds
$s^H=\tilde{s}^Q=0$ and thus $s_\text{dc}(T=0,\mu_C)=0$.
It is evident that the definition of $B_\text{dc}(T,\mu_C):=P_H(T,\mu_C,\mu_{B,\chi})$  entangles
vBag with the hadron EoS (labeled H). 
In this very aspect the finite temperature and isospin asymmetry extension of vBag($T,\mu_C$) differs from the $T=0$ case vBag discussed in \citet{Klaehn:2015}.
While the procedure does not induce a deconfinement baryon density,
i.e., $n_B^Q=\tilde{n}_B^Q= \sum_{f=u,d} \mathcal{B}_f n_f = \frac{1}{3}\left(n_u+n_d\right)$,
one has to take corrections to the energy density into account,
\begin{eqnarray}
\varepsilon^Q(T,\mu_C,\mu_B) &=&\tilde{\varepsilon}^Q(T,\mu_C,\mu_B) - B_\text{dc}(T,\mu_C) \nonumber\\
&+& T s_\text{dc}(T,\mu_C) + \mu_C n_{C,\text{dc}}(T,\mu_C)~,
\label{eq:ener}
\end{eqnarray}
with $\tilde{\varepsilon}^Q=\sum_{f=u,d} \varepsilon_f$.
The correction terms we name as the deconfinement energy density, $\varepsilon_\text{dc}(T,\mu_C)=T s_\text{dc}(T,\mu_C) + \mu_C n_{C,\text{dc}}(T,\mu_C)$. 

\section{Phase Diagram}
\label{phase}

\begin{figure}
\subfigure[~$T-\mu_B$ plane]{
\includegraphics[width=0.47\textwidth]{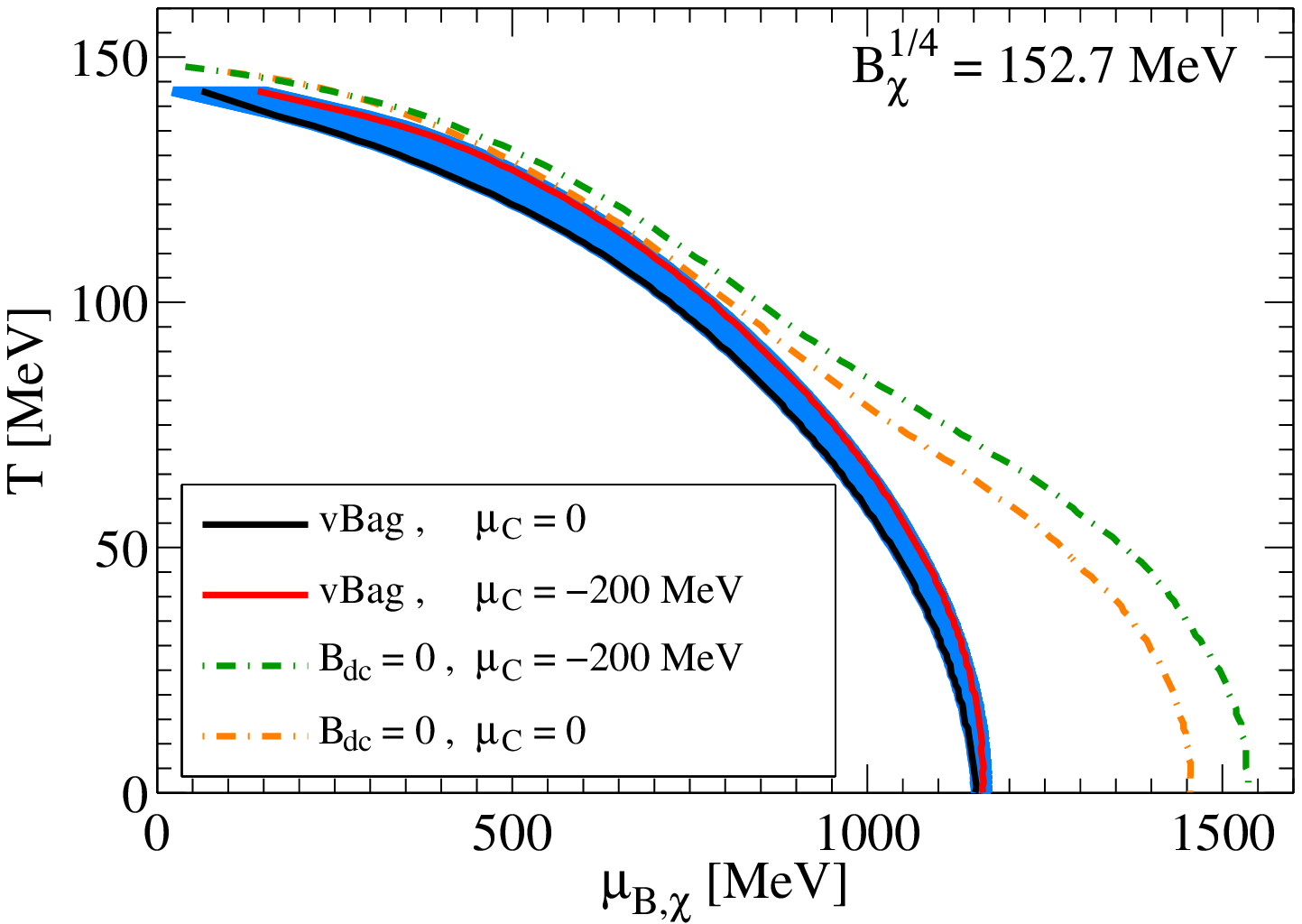}}
\subfigure[~$T-n_B$ plane]{
\includegraphics[width=0.48\textwidth]{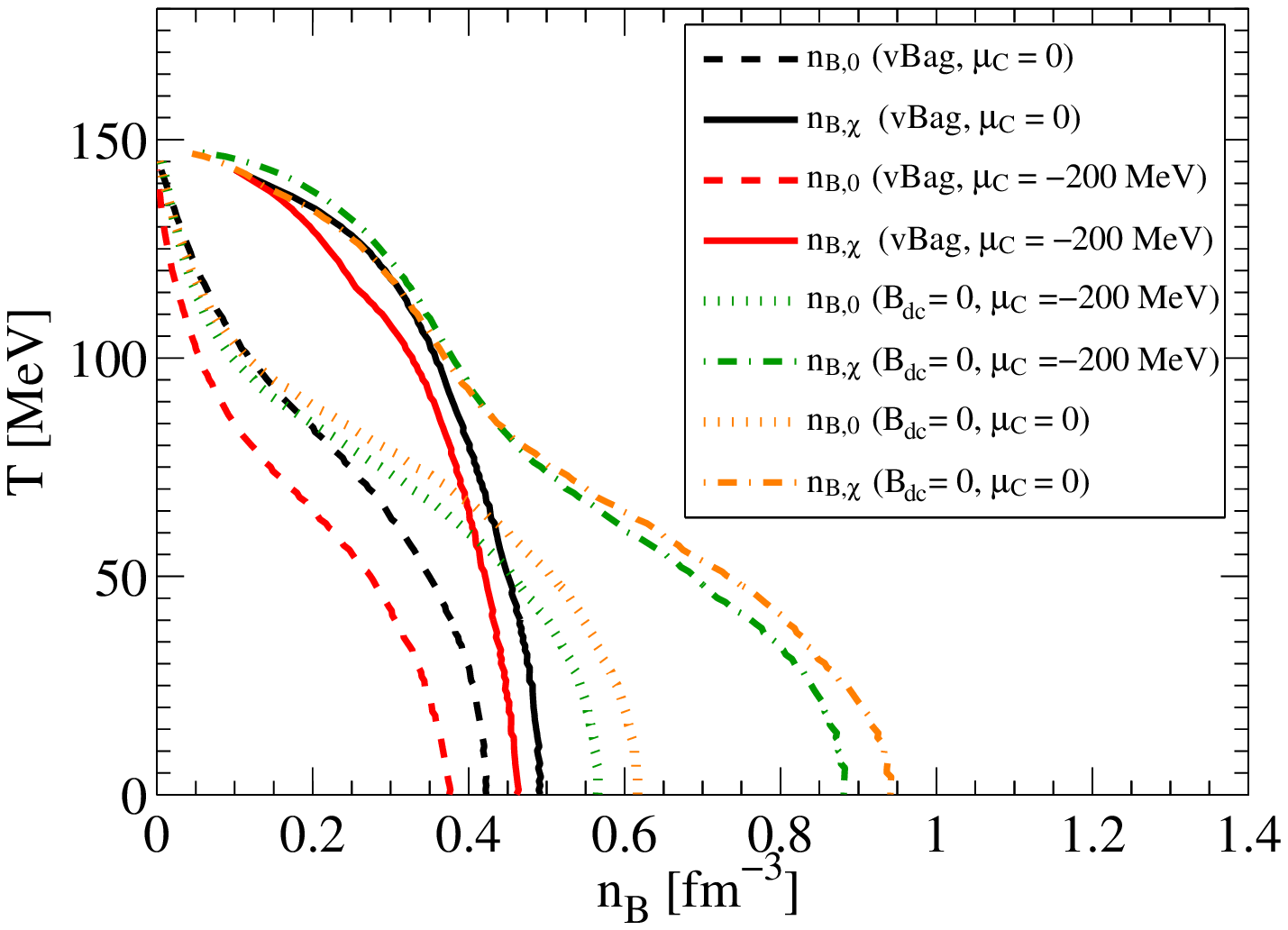}}
\caption{(color online) {\em Upper panel:} Phase diagram for the transition from HS(DD2) to vBag
with a selected single flavor chiral bag constant, $B_{\chi}^{1/4}=152.7$~MeV for
$\mu_C=0$ (symmetric matter) and $\mu_C=- 200$ MeV (strongly asymmetric matter)
comparing our and the `standard' approach ($B^{\rm dc}=0$).
{\em Bottom panel:} Corresponding density dependence of the phase diagram (see text for definitions).}
\label{fig2}
\end{figure}
\begin{figure}[htp!]
\includegraphics[width=0.48\textwidth]{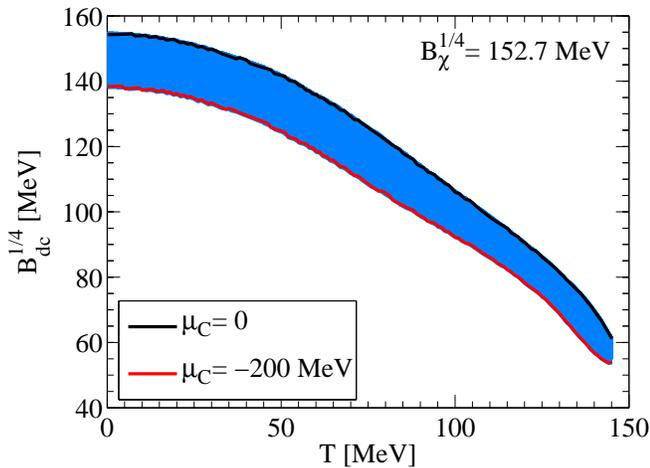}
\caption{(color online) Temperature dependence of the deconfinement bag constant for $\mu_C=0$,  $\mu_C=-200$ MeV, and intermediate values (blue band).}
\label{fig3}
\end{figure}

Fig.~\ref{fig1} illustrates the difference of our approach to the `standard' phase transition construction. 
We choose a point in the phase diagram ($\mu_B=1125$ MeV, $\mu_C=-150$ MeV, $T=30$ MeV)
and examine the pressure dependence to changes of one of these variables.
In general the vector coupling is set to zero except for set (a) where we show two cases, $K_v$=0 and $K_v=2\times 10^{-6}$ MeV$^{-2}$.
Note, that only quarks, neutrons and protons are considered, leptons, further baryons
and purely thermal excitation states, e.g., photons, pions, and gluons are not taken into account.
For the purpose of illustration we chose the nuclear HS(DD2) EoS from \citet{Hempel:2009mc} 
based on the RMF parametrization DD2 from \citet{Typel:2009sy} as it
covers finite baryon number density, temperature and isospin asymmetry.
It is available at the CompOSE database.\footnote{http://compose.obspm.fr \citep[for details, see][]{Typel:2013rza}}

\begin{figure}[t!]
\includegraphics[width=0.47\textwidth]{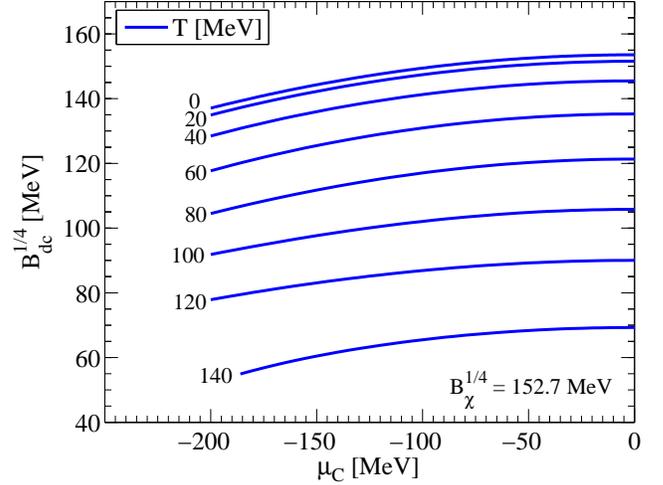}
\caption{(color online) Deconfinement  bag constant w.r.t. temperature $T$ and charge chemical potential $\mu_C$.}
\label{fig4}
\end{figure}

We refer to the `standard' phase transition construction as a prescription
that takes the EoS of both phases as independent input and favors the phase with higher pressure at
given chemical potential.
Such a model is represented by the green line in Fig.~\ref{fig1}, where the light quark single flavor chiral bag constant has
a value of $B_\chi^{1/4}=152.7$~MeV (see Table 1 in \citet{Klaehn:2015}) while we set $B_\text{dc}\equiv 0$.
With this setup we reproduce the underlying NJL model at zero temperature.
At small temperatures the model predicts the restoration of chiral symmetry at $\mu_{B,\chi}\approx1150$~MeV
defined by the condition $P^Q(\mu_{B,\chi},\mu_C,T)=0$.
However,  deconfinement as the transition from nuclear to quark matter occurs at higher chemical potential (as $P^H(\mu_{B,\text{dc}})=P^Q(\mu_{B,\text{dc}})>0$),
in this particular case at  $\mu_{B,\text{dc}}\approx 1400$~MeV. 
For all $\mu_B<\mu_{B,\text{dc}}$ HS(DD2) is determined as the thermodynamically favored EoS, 
although the quark matter model predicts the restoration of chiral symmetry on the quark level already at a significantly lower value.
The restoration of chiral symmetry is by construction not accounted for by HS(DD2).

In our approach we mimic confinement by introducing $B_\text{dc}(T,\mu_C)=P^H(T,\mu_C,\mu_B=\mu_{B,\chi})$ 
as a positive shift of the quark pressure 
as discussed in the previous section, and therefore
{\it (i)}  lower the onset of deconfinement $\mu_{B,\text{dc}}\to\mu_{B,\chi}$, and 
{\it (ii)} do not favor the nuclear EoS in a region where the dressed nucleon mass due to D$\chi$SB should be 
lower than HS(DD2) accounts for.
The result of our approach is illustrated by the solid blue line in Fig.~\ref{fig1}.
As we wish to study the consequences of our approach for the QCD phase transition in the phase diagram
we ignore the influence of vector coupling dependent terms in our model.
The reason is easily understood from the red dashed line in the same figure.
Here we chose a typical value for the vector coupling constant, $K_v=2\times 10^{-6}$ MeV$^{-2}$.
Although the thermodynamic correction terms induced by the vector interaction do visibly play an important role
at higher densities they have an almost negligible influence on the position of  $\mu_{B,\chi}$ 
and the thermodynamical quantities at this value.

\begin{figure}[t!]
\resizebox{0.45\textwidth}{!}{%
\includegraphics[width=0.45\textwidth]{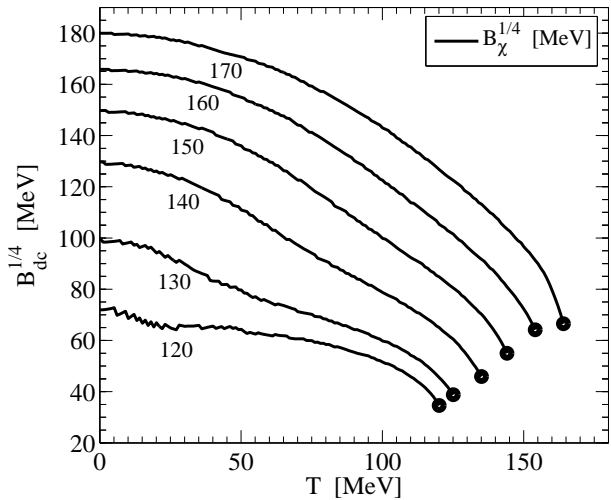}}
\caption{Temperature dependence of the deconfinement bag constant for varying chiral bag constants $B_\chi$ in symmetric matter ($\mu_C=0$). }
\label{fig6}
\end{figure}
\begin{figure*}[ht!]
\vspace{-5mm}
\subfigure[\large $-R_\varepsilon$]{
\includegraphics[width=0.46\textwidth]{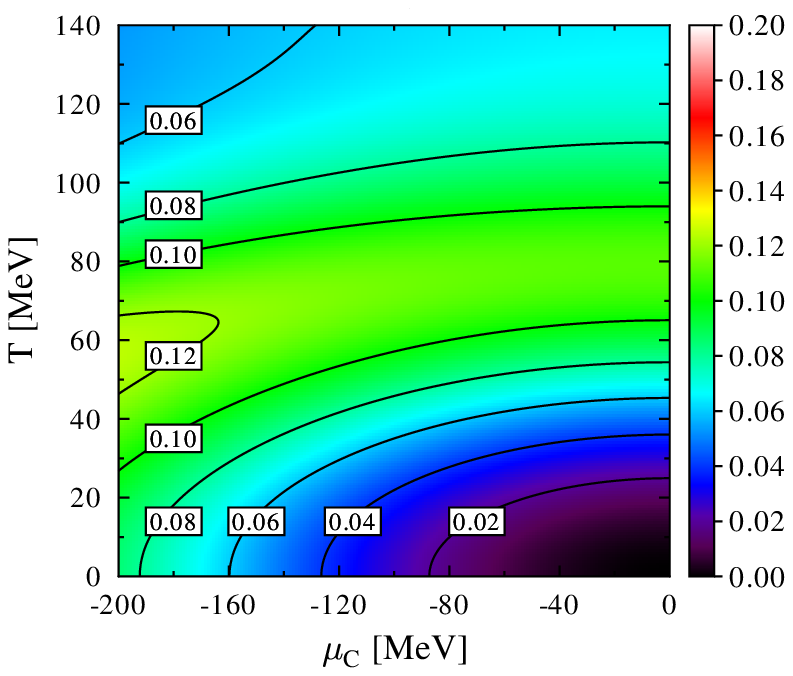}}
\hfill
\subfigure[\large $-R_s$]{
\includegraphics[width=0.46\textwidth]{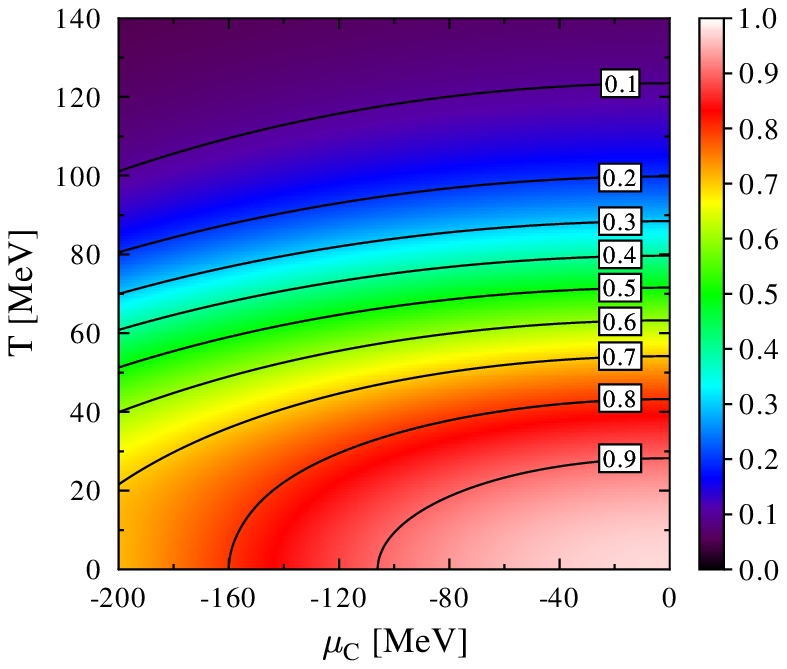}}
\vspace{-5mm}
\\
\subfigure[\large $Y_{C,\rm{dc}}$]{
\includegraphics[width=0.46\textwidth]{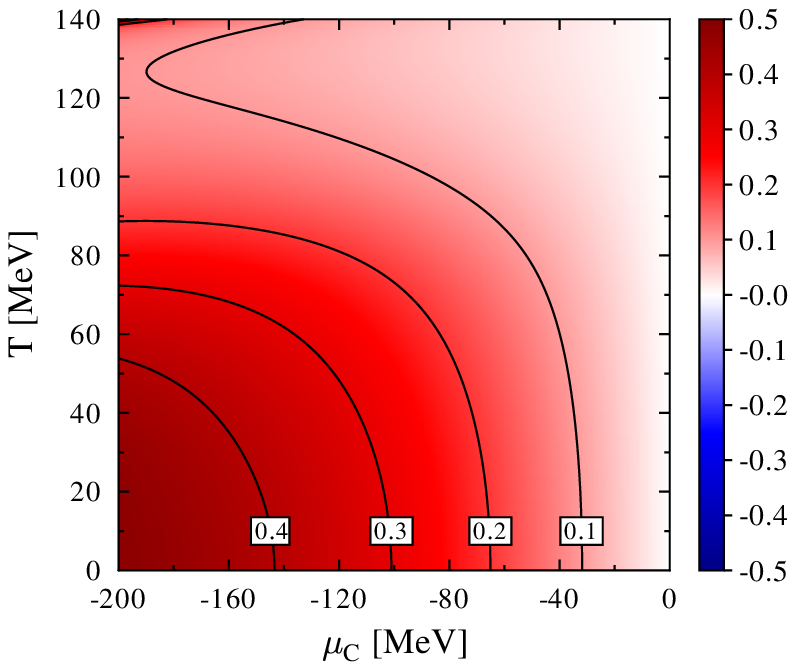}}
\hfill
\subfigure[\large $Y_C^Q$]{
\includegraphics[width=0.46\textwidth]{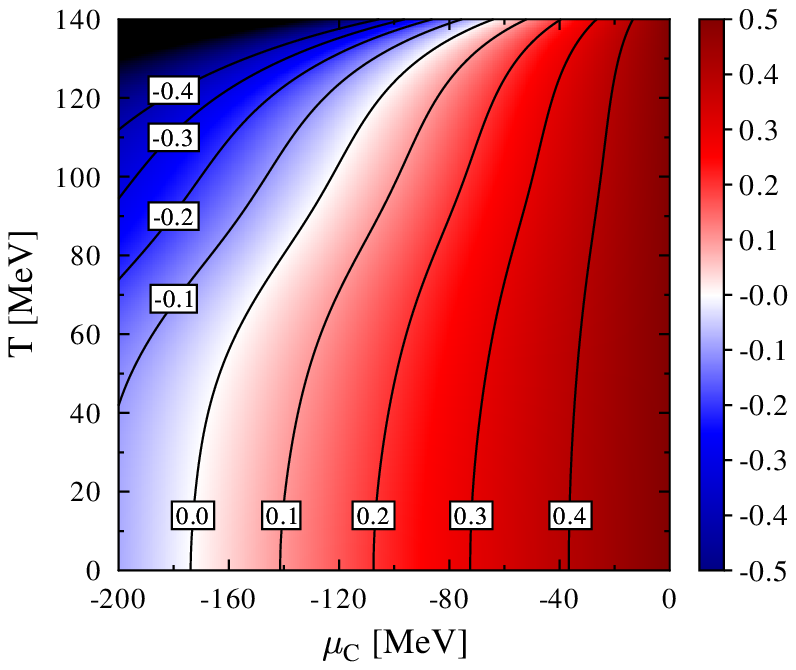}}
\vspace{-0mm}
\caption{(color online) Contour plot of the correction term to the energy density (a), entropy (b), and charge fraction (c) w.r.t. $T$ and $\mu_C$ at the phase transition (see text for definitions). (d) shows the corresponding total charge fraction of the quark phase. Notice that negative values of $R_s$ and $R_\varepsilon$ are shown. $B_{\chi}^{1/4}=152.7$~MeV.}
\label{fig7}
\end{figure*}

The lower left panel of Fig.~\ref{fig1} translates the chemical potential from the upper left panel into corresponding baryon number densities.
Both, the onset baryon densities of the phase transition ($n_{B,0}$) and of pure quark matter  ($n_{B,\chi}$) are shifted to smaller densities,
and the phase coexistence region $[n_{B,0},n_{B,\chi}]$  shrinks in comparison to the standard approach.
Our construction implies that $n_{B,\chi}$ depends only on the chiral quark model
($n_{B,\chi}$ is the slope of $P^Q(\mu_B)$ with respect to $\mu_B$ at $\mu_{B,\chi}$. $B_\text{dc}$ does not depend on $\mu_B$ and 
consequently does not contribute to $n_B$),
while $n_{B,0}$ is sensitive to both, the nuclear and quark matter EoS, as $\mu_{B,\text{dc}}=\mu_{B,\chi}$ and  $n_{B,0}=n_B^H(\mu_{B,\text{dc}})$.
We consider the effect as dramatic. While the `standard'  approach predicts an onset of the phase
transition at about  $3.5\,n_S$, our approach 
shifts the onset close to $2.5\,n_S$ (the saturation density for HS(DD2) is $n_S=0.155$ fm$^{-3}$).
Further, the density range covered by the phase coexistence region 
(visible as a plateau in the lower panel of Fig.~\ref{fig1}) is reduced from $2\,n_S$ to less than $0.5\,n_S$.
Qualitatively we find this shift of the phase transition towards smaller values for all thermodynamic degrees of freedom ($\mu_B$, $\mu_C$, and T)
together with a narrowing of the transition region in terms of the conjugated quantities ($n_B^Q$, $n_C^Q$, $s^Q$).
This is illustrated in the remaining panels of Fig.~\ref{fig1}.
While the phase transition along the temperature axis  (b) behaves qualitatively similar to
what we observe in dependence of the baryochemical potential (a) we notice a sequence of
two phase transitions (quark to hadron (I) and hadron to quark (II)) along the axis of the electronic chemical potential in column (c).

We further illustrate the differences of our and the `standard' approach by showing the phase boundary at finite temperatures and chemical potentials for 
two selected values of the charge chemical potential ($\mu_C = (-200,0)$ MeV) in Fig.~\ref{fig2}. 
As to be expected $\mu_{B,\chi}$ reduces with increasing temperature.
Moreover, our approach suggests a widening of the phase transition region in terms of density with increasing
temperature while the `standard' approach describes a narrowing.
This is related to the different temperature behavior of HS(DD2) and vBag, with a faster softening for the hadronic EoS.
In our approach the upper border of the instability region (solid black and red lines in panel (b)) is exclusively defined by the chiral transition
line predicted by vBag. In the standard approach, the restoration of chiral symmetry
is explicitly ignored. Unlike to vBag the upper border of the transition region is influenced by
the underlying nuclear matter EoS which together with the quark EoS determines the location of the phase transition.
In comparison to the differences due to a different modeling of the phase transition only small variations 
in terms of $\mu_{B,\chi}$ and $\rho_\chi$ are found for varying  $K_v$ (see Fig.~\ref{fig1}) or the isospin asymmetry (Fig.~\ref{fig2}). 

The significant changes of location and width of the phase transition region
are a direct result of the introduction of the medium dependent deconfinement bag constant $B_{\rm dc}$.
As stated earlier it maps the nuclear EoS 
at $\mu_{B,\chi}$ in order to align chiral and deconfinement phase transition.
The critical chemical potential for the chiral transition depends on the
particular value for the chiral bag constant $B_\chi$, the temperature $T$
and the isospin asymmetry.
In Figs.~\ref{fig3} and \ref{fig4} we illustrate the dependence of $B_{\rm dc}$ on temperature and charge chemical potential.
For any choice of $B_\chi$ (we discuss the model dependence on $B_\chi$ in Sec.~\ref{limits}), $B_{\rm dc}$ decreases with increasing temperature.
Simultaneously the critical baryo-chemical potential for the phase transition decreases.
The related critical temperature where the phase transition line hits the temperature axis
is indicated by filled circles in Fig.~\ref{fig6}, which also shows how it depends on $B_\chi$.

By construction, $B_{\rm dc}$ corresponds to the pressure of the hadronic and
the quark phase at the phase transition. Therefore, Figs.~\ref{fig3},
\ref{fig4}, and \ref{fig6} can also be interpreted as pressure-temperature, 
respectively pressure-charge chemical potential phase diagrams. The fact that
$B_{\rm dc}$ is decreasing with temperature implies
$\left.dP/dT\right|_{PT,\mu_C}<0$ which is opposite to the behavior in the 
nuclear liquid-gas phase transition, sometimes called an 
``entropic'' phase transition
\citep[cf.][and references therein]{hempel13,iosilevskiy14,iosilevskiy15}. 
This property could lead to interesting consequences for neutron stars and core 
collapse supernovae~\citep{yudin15,hempel15}.

According to our interpretation $B_{\rm dc}$
is a measure of the binding energy of quarks in confined hadrons.
The decreasing of this quantity with increasing temperature
indicates nicely that the system indeed becomes less bound.
Of course, the Maxwell-like approach and resulting first order transition are not an appropriate choice 
in this domain where a cross-over transition is expected.
This happens in vicinity to the critical temperature
$150<T_c<170$ MeV~\citep[cf.][]{Katz:2012JHEP,Laermann:2012PRD,Laermann:2012PRL,Katz:2014PhLB}. 
Our model is not designed to describe a cross-over phase
transition and for this reason should not be applied
at temperatures where one would expect the critical end point
and beyond.
However, the drastic decrease of $B_{\rm dc}$ with 
increasing temperature (it is rather insensitive
to the isospin asymmetry as we
illustrate in Fig.~\ref{fig4}) confirms
our interpretation of $B_{\rm dc}$ as confining
binding energy.

Although derivative terms due to the  isospin asymmetry and temperature dependence of $B_\text{dc}$, Eqs.~\eqref{eq:dBdmuc} and \eqref{eq:dBdT} respectively, 
play no role in determining the phase transition surface (in terms of $\mu_B$, $\mu_C$, and $T$), they enter the EoS as
they contribute to charge density \eqref{eq:nC}, entropy \eqref{eq:s} and energy density \eqref{eq:ener}. 
For the total energy density of quark matter we introduced $ \varepsilon^Q =  \tilde\varepsilon^Q + B_{\rm dc} + \varepsilon_{\rm dc}$
where  $ \varepsilon_{\rm dc}$ accounts for all medium dependent corrections
due to derivative terms of  $B_{\rm dc}$.
The resulting relative deviation on the phase transition hypersurface we define as $R_\varepsilon = \varepsilon_\text{dc}/\varepsilon^Q$. 
These correction terms become smaller when going to higher values of $\mu_B>\mu_{B,\chi}$, and go to zero for $\mu_B\rightarrow \infty $;  therefore $R_\varepsilon$ represents the maximal contribution of the correction terms to the energy density.
The upper left panel of Fig.~\ref{fig7} illustrates that it is always negative and that the magnitude of these corrections  is relatively small
compared to the dominant direct contribution of $B_{\rm dc}$ to $\varepsilon^Q$.
As expected, at low temperatures and small isospin asymmetry $R_\varepsilon$ turns neglible small. 
Even with increasing temperature and isospin asymmetry it remains small with values up to a magnitude of $10\%$.
However, the opposite is true for derivative corrections to the entropy and charge fraction.
The top right panel of Fig.~\ref{fig7} shows the relative contribution of the derivative term of $B_{\rm dc}$ with respect to $T$
to the total entropy density, $R_s=s_{\rm dc}/s^Q$, which is always negative.
While this quantity can reach a value of minus one at small temperatures, thus reducing the total entropy by a factor of one,
one has to keep in mind that at low temperature the entropy is small and hence this is a large relative modification of
a quantity with small value.
However, at intermediate temperatures the changes of the entropy due to the medium dependence of $B_{\rm dc}$ are
far from being negligible.
The situation turns even more extreme for the charge density $n_C^Q=\tilde n_C^Q+n_{C,\rm{dc}}$ or the corresponding
charge fractions $Y_C^Q=n_C^Q/n_B^Q=\tilde Y_C^Q + Y_{C,{\rm dc}}$.
The bottom left panel of Fig.~\ref{fig7} illustrates that derivate corrections vanish for values of $\mu_C$ close to zero.
However, with more negative values of $\mu_C$ the correction term $Y_{C, {\rm dc}}$ can reach values of up to 0.5.
For the total charge fraction (bottom right panel) this results in significant changes, generally shifting the charge fraction towards more positive values.

\section{Protoneutron Stars}
\label{pns}
We develop the presented model to investigate possible signatures of the QCD phase transition during simulations of core-collapse supernovae~\citep[cf.][]{Sagert:2008ka,Nakazato:2008su,Fischer:2011}.
Such transition is expected at high baryon densities, $n_B>n_S$, in the core of a newly born protoneutron star. The latter originates from a core-collapse supernova, being triggered from the initially imploding stellar core of a massive star leading finally to the ejection of the stellar mantle, known as {\em supernova problem}.

\begin{figure}[t!]
\resizebox{0.48\textwidth}{!}{%
\includegraphics[width=0.48\textwidth]{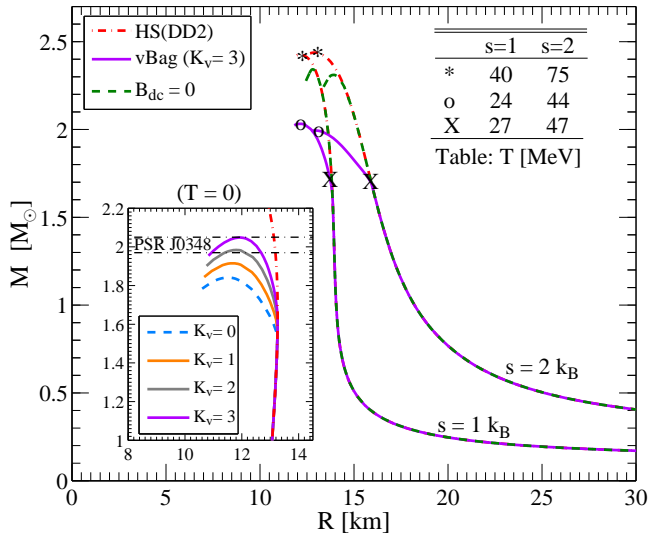}}
\caption{Mass radius relations for constant entropy per particle (s=1,2 $k_B$) {\bf for $B_\chi^{1/4}$=152.7 MeV}. The temperature of the most massive NS without quark core (*) is in general much larger than the corresponding value for a NS with quark core (o). By construction, the temperature in the transition point (X) for hadron and mixed phase is equal.}
\label{figPNS}
\end{figure}

\begin{figure}[t!]
\resizebox{0.48\textwidth}{!}{%
\includegraphics[width=0.5\textwidth]{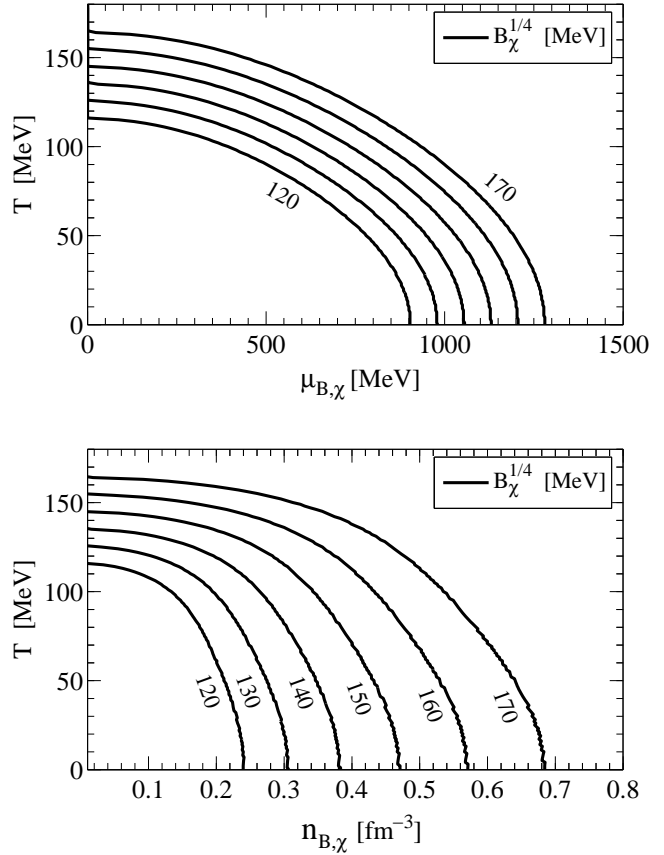}}
\caption{{\em Upper panel:} Symmetric matter phase diagram for varying chiral bag constants $B_\chi$. 
{\em Bottom panel:} Corresponding onset density for pure quark matter.}
\label{fig5}
\end{figure}

We first adjust our model such that a cold neutron star can form a quark matter core.
The electron gas component in $\beta$-equilibrium is included under the assumption of global charge neutrality. In such a case they represent a uniform background, which does not influence the phase transition.
As pointed out in \citet{Klaehn:2015} this requires repulsive vector interactions as provided by vBag in order to
describe a maximum neutron star mass of about 2~M$_\odot$ which is in agreement with the results of 
\citet{Antoniadis:2013} and \citet{Fonseca:2016tux}.
The inset of Fig.~\ref{figPNS} shows that vBag provides such massive neutron stars for a vector coupling constant of $K_v>2\times 10^{-6}$ MeV$^{-2}$.
In order to mimic a typical temperature profile for ''hot'' protoneutron stars we assume an isentropic EoS at two different values
for the entropy per particle (s=1,2 k$_{\rm B}$).
In  Fig.~\ref{figPNS} we show the resulting mass-radius relations and indicate the core temperature at three characteristic points.
In particular one observes that the quark matter core tends to be significantly cooler than the corresponding core of a purely hadronic star.
This is a direct result of the entropic phase transition to quark matter which has a smaller value of $s$ at any given point in the phase diagram, compared to the nuclear EoS.
{\bf We further point out, that the chosen chiral bag constant of   $B_\chi^{1/4}$=152.7 MeV results in a transition density of approximately three times saturation
density in symmetric matter, cf. Fig.\ref{fig5} (lower panel). 
Although the model allows different choices and correspondingly different transition densities we consider this value as
an acceptable estimate for the lower limit on a possible phase transition to quark matter.}

The study of possible consequences of the quark-hadron phase transition based on vBag, for the supernova dynamics as well as on potential observables, e.g., the neutrino and gravitational wave signals, extends beyond the scopes of the present paper. This will be explored in an upcoming article.

\section{Model limitations}
\label{limits}
We motivated that the deconfinement bag constant has a temperature and isospin dependence which is
implicitly given via the hadronic EoS.
The chiral condensate and therefore the bag constant $B_\chi$ 
should show an explicit temperature dependence as well. 
We neglect such dependencies and assume the same fixed value for $B_\chi$ at all $T$
and study the impact of varying conditions for the chiral transition to
construct our phase diagrams. 
Fig.~\ref{fig5} illustrates that with decreasing $B_\chi$ also $\mu_{B,\chi}$ and $n_{B, \chi}$ will reduce substantially.
It is formally not difficult to rewrite our approach for a fully medium dependent chiral bag constant $B_\chi$.
However, modeling a {\it meaningful} medium dependence 
requires a more elaborate approach than our phenomenological model. 
We therefore leave this important issue for future investigations
and rather provide a brief discussion of the validity of our approach.
Before we proceed we emphasize that we understand vBag as an extension of the thermodynamic bag model
rather than a NJL model parameterization even though the extensions are motivated from a NJL model perspective.
In this sense all limitations we will discuss in the following hold in a similar way for vBag {\it and} tdBAG.

\begin{figure}[t!]
\resizebox{0.48\textwidth}{!}{%
\includegraphics[width=0.5\textwidth]{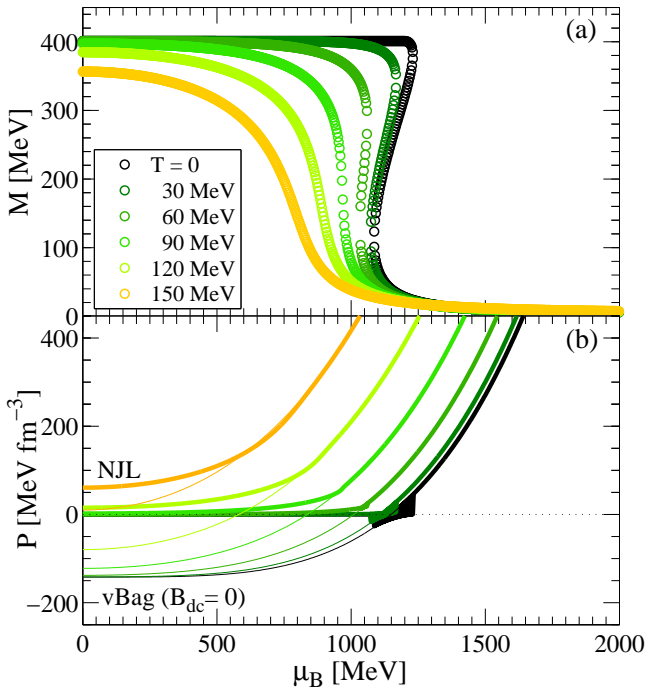}}
\caption{(a) Mass gap solutions of the  NJL model used in \citet{Klaehn:2015} at different temperatures. 
For temperatures smaller than 90 MeV  multiple mass gap solutions exist in vicinity of the critical quark chemical potential $\mu_\chi$. 
(b) Pressure of symmetric two flavor quark matter obtained in the NJL model compared to the ideal gas pressure with bare quark masses at the same temperatures. 
The ideal gas pressure has been shifted by a temperature dependent offset to reproduce the high density behavior of the NJL results (details in text).
For temperatures below 90 MeV this shift is nearly constant.
}
\label{figMmuDiffT}
\end{figure}

In \citet{Klaehn:2015} we illustrated that vBag at zero temperature can be parameterized to reproduce
the EoS of a NJL model including the typical first order phase transition in good approximation. 
Fig.~\ref{figMmuDiffT}(a) illustrates that at temperatures larger than 90 MeV only one mass gap solution exists
at any given quark chemical potential and thus no distinct first order phase transition is obtained.
For temperatures beyond this value one additionally observes a decrease of the mass gap already at zero
quark chemical potential.

\begin{figure}[t!]
\resizebox{0.49\textwidth}{!}{%
\includegraphics[width=0.49\textwidth]{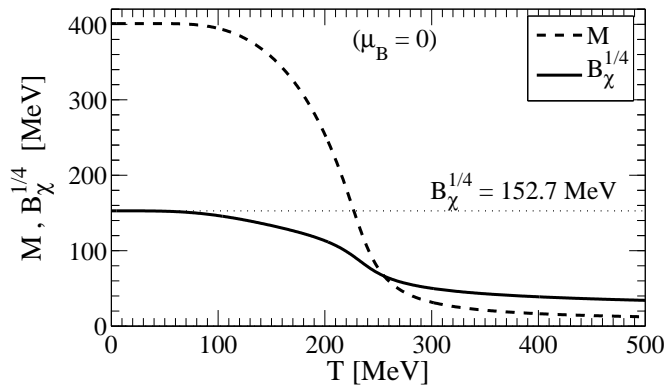}}
\caption{Temperature dependence of the mass gap solutions of the NJL model used in \citet{Klaehn:2015} at zero quark chemical potential and the resulting 
one flavor chiral bag constant.}
\label{figMBT}
\end{figure}
This behaviour of the dressed mass $M$ at $\mu=0$ is shown for a wider range of temperatures in Fig.~\ref{figMBT}. 
Additionally we plot the difference of the pressure of light quarks with temperature dependent dressed mass to the pressure of bare mass quarks at $\mu=0$. In both cases the vacuum pressure according to Eq.~(17) in \cite{Klaehn:2015} is included. The temperature dependence up to 100 MeV is negligible.

Fig.~\ref{figMmuDiffT}(b) shows the full dependence of the pressure on the chemical potential for different temperatures as we obtain it within the NJL model considering only the scalar contributions to the self energy (thick lines). The increasing offset at $\mu=0$ with increasing temperature results from the thermal excitation of quarks and antiquarks. We further plot the corresponding results for an ideal gas of quarks with undressed bare mass (thin lines). This corresponds to the vBag model with $K_v=0$. The offset between NJL and vBag results is determined at zero $\mu$ in the same way we determine  $B_\chi$, however for this plot at finite temperature. Fig.~\ref{figMBT} shows this quantity. Again, the difference to the $T=0$ result is small for temperatures below about 90 MeV. Moreover, Fig.~\ref{figMmuDiffT}(b) shows that at sufficiently large chemical potential both approaches give the same result at all temperatures. Of course, this is the same behavior one observes at $T=0$ and which initially motivated us to introduce vBag as a simplified model EoS that does not require to solve gap equations explicitly.

In conclusion, although not intended to fully mimic NJL model calculations vBag sufficiently reproduces the NJL results for temperatures up to 90 MeV. 
Higher temperatures already correspond to conditions which are typically out of reach for astrophysical applications, e.g., core-collapse supernovae and neutron star scenarios. Finally, we remark that neither NJL nor vBag nor the hadronic EoS HS(DD2) are developed for large temperatures and small chemical potentials, e.g. to study the QCD phase transition as in particular in the transition domain thermally excited mesons (e.g. pions) and additional nuclear resonances have to be taken into account.

Moreover, neither strange quarks nor hyperons are considered in this study. It is a worthwhile question to investigate whether the transition to quark matter happens before or after hyperon degrees are excited. However, as little is known about hyperon interactions we once again  would have to make model assumptions which crucially affect the answer to this question, cf.~\citet{Bonanno:2011ch} and ~\citet{Masuda:2015kha}.

\section{Summary and Discussions}
\label{summary}
In this article we introduced an important extension of the recently introduced quark matter EOS vBag~\cite[for details, see][]{Klaehn:2015} to finite temperature and arbitrary isospin asymmetry. 
As an extension of tdBag it takes  D$\chi$SB explicitly into account in terms of the chiral bag constant $B_\chi$.
It imposes the restoration of chiral symmetry at an associated critical chemical potential ($\mu_{B,\chi}$). 
This first model parameter is complemented by a second one, $K_v$, which is the vector coupling constant due to repulsive vector interactions -- not included in tdBag -- which stiffens the EoS with increasing density. 
The last main characteristics of vBag is the inherent connection to a given hadronic EoS in terms of the medium-dependent
deconfinement bag constant ($B_{\rm dc}$).
It is determined entirely by the chiral transition and a selected hadronic EoS -- and in this sense is not a free parameter of vBag. 
With this procedure we assure that chiral and deconfinement phase transitions coincide, assuming that the transition is of first order. 
This purely phenomenologically introduced medium dependence of the deconfinement bag constant of vBag to the hadronic EoS
bases on the assumption of simultaneous chiral symmetry restoration and deconfinement. In this sense, our approach differs from quark models
that aim to determine the medium dependence of a bag constant microscopically, ~\citep[cf.][and references therein for a discussion of NJL and DS analyses respectively]{Buballa:2003qv,Roberts:2000aa}.

With vBag and the extension to finite temperatures and isospin asymmetry we 
provide a novel quark-matter EoS for a broad spectrum of applications in astrophysics, e.g., simulations of neutron star mergers, core collapse supernovae and protoneutron star cooling, where commonly the simple tdBag-model has been employed~\citep[cf.][and references therein]{Pons:2001ar,Sagert:2008ka,Nakazato:2008su,Fischer:2011}. 
tdBag does not account for chiral physics and without perturbative corrections violates the 2~M$_\odot$ neutron star mass constraint due to missing repulsive interactions. 
Here, vBag provides solutions that cure both problems without losing the simplicity and flexibility of tdBag. 

A natural consequence of the extension to finite temperatures and isospin asymmetry is the emergence of an implicit temperature and isospin asymmetry dependence of the deconfinement bag constant, $B_\text{dc}(T,\mu_C)$. It is determined by the pressure of the hadronic EoS at the chiral transition. In addition, corrections to entropy and charge density arise from a thermodynamically consistent treatment. 
The phase transition surface in the phase diagram (in terms of $\mu_B$, $\mu_C$, and $T$) is independent from these correction terms.
Furthermore, the resulting corrections to the energy density are small while we find significant contributions
to the entropy density and charge fraction of quark matter.
Overall we find that a simultaneous chiral symmetry restoration and deconfinement make the quark phase more similar to the hadronic phase 
regarding its temperature and asymmetry dependence.

The hybrid EoS is constructed based on our choice of the nuclear RMF EoS HS(DD2).
Starting from the assumption $\mu_{B,\text{dc}}=\mu_{B,\chi}$ 
the phase boundary in the phase diagram shows distinct differences to the common approach which is used in astrophysics.
The standard procedure results in significantly higher transition densities than our approach with finite $B_\text{dc}$.
Moreover, we find a different behavior of the phase boundary towards increasing temperatures. 
The coexistence region between the onset of quark matter and pure quark matter broadens, 
unlike to the `standard' approach.
Since the deconfinement bag constant maps the pressure of the hadronic EoS at the transition line, 
its medium dependence is strongly affected by the hadronic EoS. 
It decreases with increasing temperature and rather weakly depends on  the isospin asymmetry. 
Furthermore, we briefly discussed the expected impact of a temperature dependent chiral bag constant by choosing different values of $B_\chi$.
The phase boundaries shift to lower(higher) densities for lower(larger) values of $B_\chi$. 
This important aspect will be further explored in future studies, as it is beyond the scope of the presented article.

At temperatures in excess of about 100~MeV we are leaving the domain of validity of our purely  nucleonic EoS; nuclear resonances and mesons such as the pion are not explicitly included in the description of the hadronic medium of HS(DD2). Nevertheless, the qualitative behavior of the vBag phase boundary seems to be in agreement with predictions of lattice QCD and data from heavy-ion collision experiments  with one important exception: 
the phase transition we describe in our approach is by construction always of 1st-order. 

\section*{Acknowledgement}
The authors acknowledge support from the Polish National Science Center (NCN) under grant numbers DEC-2011/02/A/ST2/00306 (TF) and UMO-2013/09/B/ST2/01560 (TK). MH acknowledges support from the Swiss National Science Foundation. Partial support comes from ``NewCompStar'', COST Action MP1304.

\bibliography{references}
\end{document}